\documentclass[twocolumn,prl,floatfix,a4paper]{revtex4-2}
\usepackage{blindtext}
\usepackage{comment}
\usepackage[english]{babel}
\usepackage[utf8]{inputenc}
\usepackage[T1]{fontenc}
\usepackage{cancel,xcolor}
\usepackage{caption}
\usepackage{dsfont}
\usepackage{amsmath}
\usepackage{amssymb}
\usepackage{amsfonts}
\usepackage{amsthm}
\usepackage{graphicx}
\usepackage[scale=0.8]{geometry}
\usepackage{float}

\begin{document}
\title{Critical-level wave absorption controls chaos in a wave--mean flow system}

\author{Aymeric L\'evy-Dauchez$^{1,2}$}
\author{Louis-Philippe Nadeau$^{3}$}
\author{Antoine Venaille$^{1}$}
\email{antoine.venaille@ens-lyon.fr, aymeric.levy-dauchez@lmd.ipsl.fr}

\affiliation{$^1$ CNRS, ENS de Lyon, Laboratoire de Physique UMR5672, Lyon, France,$^2$ Universit\'e PSL, \'Ecole Normale Sup\'erieure, D\'epartement de g\'eosciences, Laboratoire de M\'et\'eorologie Dynamique, Paris, France, $^{3}$ Institut des Sciences de la Mer de Rimouski, Universit\'e du Qu\'ebec \`a Rimouski, Rimouski, Qu\'ebec G5L 3A1, Canada}

\begin{abstract}
Wave–mean flow interactions in stratified fluids can drive spontaneous oscillations, yet the mechanisms controlling transitions between periodic and chaotic regimes remain poorly understood. Using a minimal one-dimensional model of wave-driven mean-flow reversals motivated by the quasi-biennial oscillation (QBO), we identify two distinct mechanisms that suppress the formation of multiple extrema in the velocity profile: critical-level momentum deposition by a broad spectrum of undamped small-scale gravity waves, and wave-amplitude growth due to exponential density stratification. Although physically distinct, both mechanisms inhibit the route to chaos and reduce sensitivity to small-scale viscosity, substantially widening the region of periodic behavior in parameter space. These results identify a general internal mechanism for the stabilization of wave-driven oscillations, with direct implications for QBO-like phenomena in geophysical and astrophysical flows.
\end{abstract}

\maketitle 

Wave--mean flow interactions in stratified fluids can drive spontaneous oscillations across a wide range of physical systems, from laboratory experiments \citep{plumb_78,seelig2015can,semin_nonlinear_2018} to planetary atmospheres and models of stellar interiors \citep{baldwin_2001,showman_atmospheric_2019}. A prominent geophysical example is the quasi-biennial oscillation (QBO) of equatorial stratospheric winds, which influences interannual variability at the planet scale \citep{anstey2022impacts}. A wide range of QBO-like regimes, including irregular and chaotic oscillations, has been reported across modeling hierarchies, from one-dimensional reduced models \citep{kim_gravity_2001,renaud_periodicity_2019} to idealized direct numerical simulations \citep{couston_order_2018,daniel2025self,bremaud2025idealized} and general circulation models of planetary atmospheres and stratified layers of stellar interiors \citep{showman_atmospheric_2019,bardet2022joint}, indicating that the mechanisms controlling the robustness of nearly-periodic mean-flow reversals remains incompletely understood.

These dynamics hinge on wave--mean flow interaction, driven by a spectrum of internal gravity waves whose parameterization in general circulation models remains challenging \citep{lindzen_holton_68,dunkerton_97,sutherland2019recent,holt2022evaluation}. We focus on the minimal configuration leading to low-frequency oscillations: vertically propagating waves generated below a vertically sheared zonal flow and absorbed within it.  Horizontally propagating waves in laterally sheared flows are not required for the oscillation itself, although they can perturb these oscillations, as in the case of recent QBO disruptions on Earth \citep{osprey_unexpected_2016}.

Early reduced models by Holton, Lindzen and Plumb explained the spontaneous emergence of oscillations by considering a standing planetary wave source \citep{holton_lindzen_72,plumb_77}. Subsequent work showed that modifying the wave spectrum or its parameterization can qualitatively change the dynamics \citep{saravanan_90,dunkerton_97}, including the emergence of oscillations \cite{campbell2005constraints1,campbell2005constraints2} or, surprisingly, the stabilization of oscillations against chaos under broadband wave forcing \citep{leard_multimodal_2020}, though the mechanisms behind these effects remain unclear.

\begin{figure}
\centering
\includegraphics[width=\columnwidth]{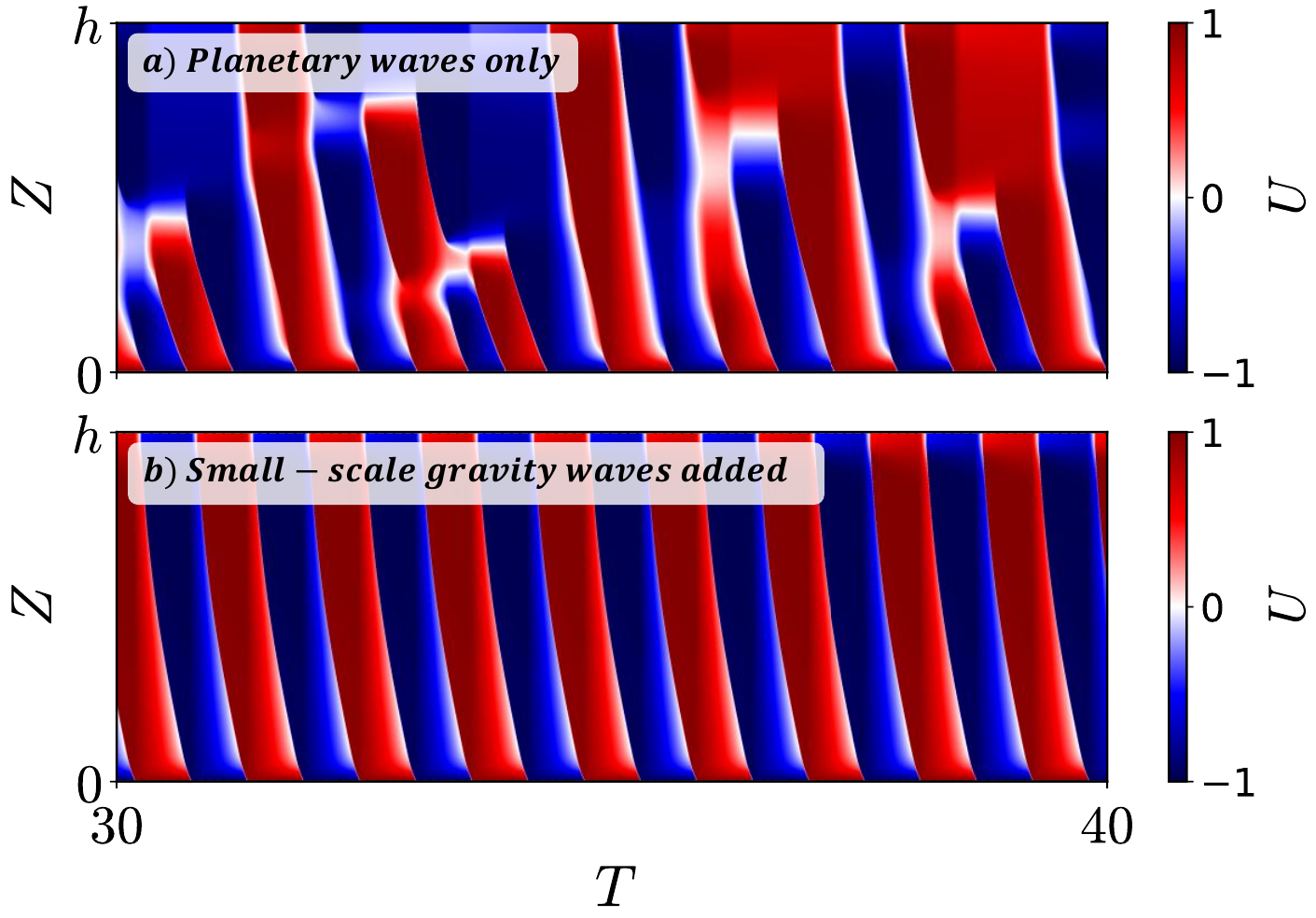}
\caption{Simulations of Holton-Lindzen-Plumb model for horizontal mean flow reversals a) forced  by a single standing planetary-scale wave (P77$\rho$ in the text, with $\alpha=0$) b) forced with an additional contribution modelled by the  Lindzen-Holton 1968 parameterization (LH68 with $\alpha=0.2$). In both cases, $h_\rho=1.4$,  $h=4$ and $Re=100$.}
\label{fig:compareParams}
\end{figure}

\begin{figure*}
\centering
\includegraphics[width=2\columnwidth]{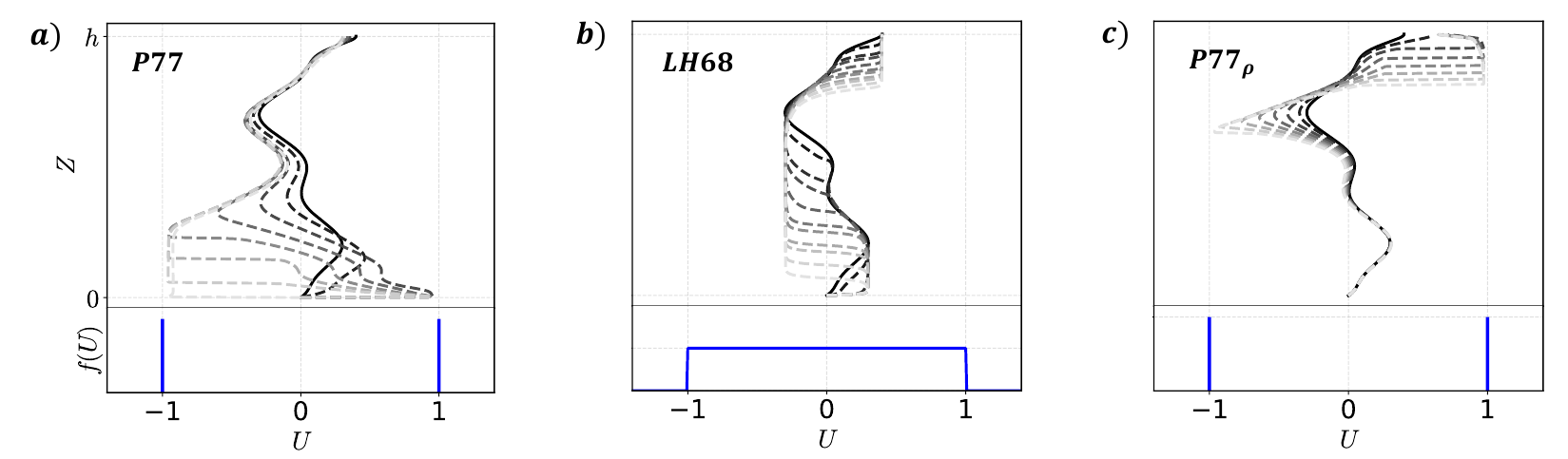}
\caption{Temporal evolution of a wind profile (initial state: plain black line) a) $(Re,\alpha,h,h_\rho)=(250,0,4,\infty)$ b) idem with  $Re\rightarrow 0$  and $\alpha Re= 250$  c) same as panel a with $h_\rho=0.4$. In dimensional units, cases (a) and (c) correspond to $F_p \neq 0$, $F_{\text{sgw}} = 0$, while case (b) corresponds to $F_p = 0$, $F_{\text{sgw}} \neq 0$. All cases share the same viscosity value.  The corresponding imposed 
momentum flux density $f(U)$ is plotted in each case. Time between initial and final profiles: $\Delta t = (0.7,\, 0.8,\, 10^{-3})$ for cases a, b, c respectively.}
\label{fig:early_evolution}
\end{figure*}
In this Letter, we isolate the mechanisms responsible for the stabilization of periodic oscillations in the presence of  broadband wave forcing, building on the finding that transitions to chaos in wave--mean flow systems are associated with the amplification of vertical modes containing multiple velocity extrema \citep{chartrand2024recovering}. Using a minimal extension of the Holton--Lindzen--Plumb framework, we show that two distinct processes suppress these multiple extrema: (i) state-dependent momentum deposition through critical-level absorption of undamped small scale gravity waves, captured by the seminal Lindzen-Holton 1968 parameterization \citep{lindzen_holton_68}; and (ii) density-driven amplification of wave forcing due to exponential stratification. Although physically distinct, both mechanisms suppress secondary shear layers and thereby inhibit the route to chaos, enlarging the periodic regime.

\textbf{Model.} We consider a stratified fluid layer of depth $H$, with an 
exponential background density profile characterized by a scale height $H_\rho$. 
The flow is forced from below by two counter-propagating planetary waves of phase 
speeds $\pm c$ and momentum fluxes $\pm F_p$, subject to radiative attenuation 
over a vertical length scale $\Lambda$, and supplemented by a broad spectrum of 
undamped small-scale gravity waves (SGW) carrying zero net momentum flux, with 
eastward and westward contributions $+F_{sgw}$ and $-F_{sgw}$, respectively. 
The system is governed by four nondimensional parameters,
\begin{equation}
(Re,\,\alpha,\,h,\,h_\rho) 
= \left(\frac{F_p\Lambda}{\nu c},\;
        \frac{F_{sgw}}{F_p},\;
        \frac{H}{\Lambda},\;
        \frac{H_\rho}{\Lambda}\right),
\label{eq:params}
\end{equation}
which control, respectively, the relative importance of wave forcing and viscosity, 
the relative amplitude of SGW forcing, the degree of planetary-wave attenuation, 
and the degree of exponential stratification.

In the quasilinear approximation, the horizontally averaged zonal flow $\bar{u}(z,t)$ 
evolves under the divergence of wave-driven momentum fluxes and viscous diffusion. 
In an anelastic medium, density stratification amplifies wave amplitudes with 
altitude, introducing an exponential factor $e^{Z/h_\rho}$ that weights the 
wave-flux divergence. Introducing nondimensional variables $U=\bar{u}/c$, 
$Z=z/\Lambda$, and $T=tF_p/(c\Lambda)$, we obtain
\begin{equation}
\partial_T U = Re^{-1}\,\partial_Z^2 U 
- e^{Z/h_\rho}\,\partial_Z\!\left[\mathcal{F}_p[U] 
+ \mathcal{F}_{sgw}[U]\right],
\label{eq:model}
\end{equation}
with $Z \in [0, h]$, no-slip at the lower boundary $U(0,T)=0$, and a 
stress-free condition $\partial_Z U=0$ at the top. Here $\mathcal{F}_p$ and $\mathcal{F}_{sgw}$ 
denote the vertical momentum fluxes associated with planetary and small-scale 
gravity waves, respectively. Both wave populations are forced exclusively at the lower boundary,
representing wave generation in the underlying troposphere. Changing the lower boundary condition for the mean-flow to free-slip does not qualitatively affect periodicity recovery (Supplementary Material), consistent with structural stability observed in previous studies on this model  \cite{rajendran2016synchronisation,renaud_holton-lindzen-plumb_2020}.

Planetary-wave forcing follows \citet{plumb_77}. The two waves undergo 
Doppler-shifted radiative attenuation, yielding the net momentum flux
\begin{equation}
\mathcal{F}_p[U] = e^{-\int_0^Z (1-U)^{-2}\,dZ'}
                 - e^{-\int_0^Z (1+U)^{-2}\,dZ'},
\label{eq:Fp}
\end{equation}
where $|U|=1$ corresponds to critical levels at which absorption becomes complete. For simplicity, we assume an altitude-independent attenuation length $\Lambda$, whereas it varies in real atmospheres (see Supplementary Material for the consequences).

\begin{figure*}
\centering
\includegraphics[width=\textwidth]{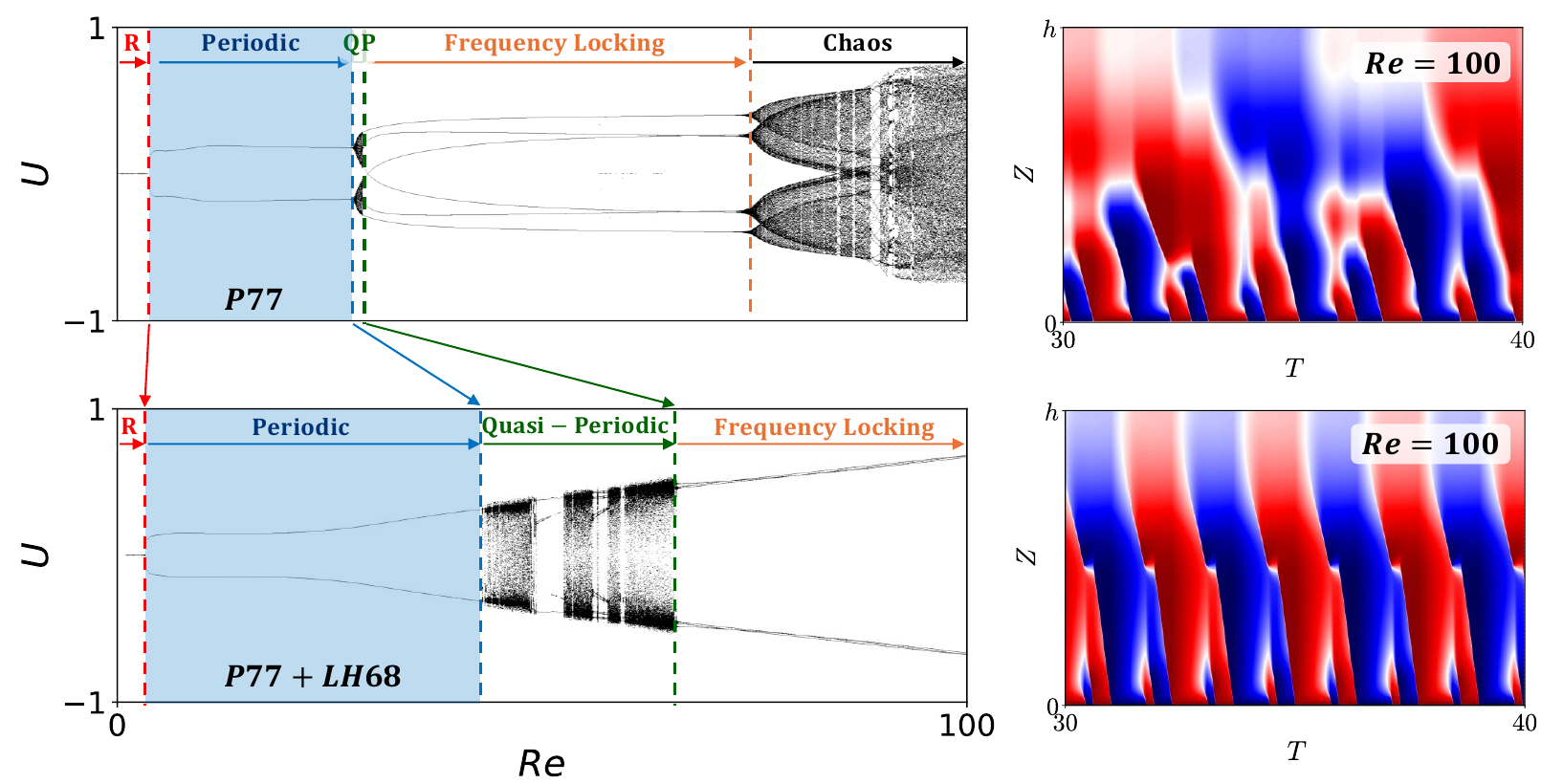}
\caption{Effect of adding LH68 parameterization to P77 model, with $h_\rho=\infty$, $h=4$. a) Bifurcation diagram comparing a situation without LH68 ($\alpha=0$) and a case with  LH68 parameterization ($\alpha=1$) b) Hovmöller diagram showing the structure of the QBO state at Re =100, with and without additional LH68 parameterization.}
\label{fig:P77LH68bif}
\end{figure*}

Small-scale gravity-wave forcing follows \citet{lindzen_holton_68} (LH68 hereafter), assuming that their attenuation length scale is much larger than the domain height. The key 
mechanism is critical-level absorption: a wave of phase speed $C$ (in units of 
$c$) deposits all its momentum at the lowest altitude where $U=C$, and carries 
none higher. Assuming a uniform distribution of phase speeds $C\in[-1,1]$ at the 
lower boundary and integrating over the spectrum yields the effective closure
\begin{equation}
\mathcal{F}_{sgw}[U](Z) = 
\begin{cases}
-\alpha\, U(Z) & \text{if } U(Z) \notin U\!\left([0,Z[\right), \\
0              & \text{otherwise},
\end{cases}
\label{eq:Fsgw}
\end{equation}
where the condition $U(Z) \notin U([0,Z[)$ encodes critical-level 
absorption: the flux carried by a given wave with phase speed $C = U(Z)$ vanishes above its lowest critical level, making $\mathcal{F}_{sgw}$ a nonlinear, state-dependent functional of $U$. We emphasize that $\mathcal{F}_{sgw}$ is not an arbitrary additional term: it is the weak-attenuation limit of a continuum of waves with different phase speeds, each obeying the P77 absorption law \citep{saravanan_90}. LH68 thus captures, more simply and cheaply than P77 (see Supplemental Material), what we expect to be the generic effect of nearly undamped gravity waves. For analytical simplicity, we use a waterbag momentum-forcing distribution, sufficient to reproduce the generic behavior found for broad wave-forcing distributions in \cite{leard_multimodal_2020}.

In the limit $\alpha=0$ and $h_\rho\to\infty$, we recover the texbook model as formulated by \citet{plumb_77} (P77 hereafter), who also discussed the case of finite $h_\rho$ (P77$\rho$ hereafter). Representative terrestrial values are $h \approx 3$, $h_\rho \approx 1$ and $\alpha \approx 0.2$ \citep{match_mean-flow_2020}, and $Re \approx 10$ yields QBO-like oscillations. However, $\alpha$ and especially $Re$ are poorly constrained in practice: owing to the uncertainty in turbulent viscosity, $Re$ spans $2$--$400$ \citep{renaud_periodicity_2019}, which in the P77 model encompasses the full range of dynamical regimes, from rest to chaos. Why, then, are nearly periodic states so prevalent in both more realistic models and observations?

\textbf{Results.} Figure \ref{fig:compareParams} illustrates the central result of this 
Letter. In the absence of small-scale gravity-wave forcing ($\alpha=0$), the P77$\rho$ model produces a chaotic mean-flow reversal pattern characterized by the irregular development of high-altitude secondary jets (panel~a). Adding the LH68 parameterization ($\alpha=0.2$) at otherwise 
identical parameter values restores a clean periodic oscillation 
(panel~b). This dramatic transition from irregular, broadband reversals 
to a well-defined QBO-like cycle motivates the mechanistic analysis 
that follows.

Figure~\ref{fig:early_evolution} compares the early evolution of a wind 
profile with multiple extrema under three configurations:  P77 alone ($\alpha=0$, 
$h_\rho\to\infty$), LH68 alone 
(no planetary waves, with $\alpha Re=250$), and P77$\rho$ ($\alpha=0$, $h_\rho=0.4$). Under 
P77, local extrema are amplified as fronts descend, with 
velocities converging toward $|U|\to 1$ \cite{plumb_77}; at large $Re$, secondary shear 
layers can grow unchecked \cite{chartrand2024recovering}. Under LH68, fronts also form and propagate 
downward, but the velocity field cannot 
develop amplitudes beyond its initial range, and existing extrema are 
progressively erased. Although multiple jets initially co-exist, the system eventually reaches a steady state with a homogenenous velocity profile above a lower boundary layer \cite{lindzen_holton_68}. Under P77$\rho$, the exponential amplification 
of wave forcing at upper levels ($e^{Z/h_\rho}$) dominates over 
radiative attenuation when $h_\rho < 1$ (i.e.\ when the density scale 
height is smaller than the wave attenuation length scale $\Lambda$), accelerating convergence 
toward $|U|\to 1$ at altitude and suppressing secondary extrema before 
they can develop. Since chaotic dynamics in the HLP model arise through 
the amplification of modes with multiple velocity extrema \citep{chartrand2024recovering}, both mechanisms are expected to inhibit this route to chaos.

Figure~\ref{fig:P77LH68bif} shows bifurcation diagrams as a function 
of $Re$ with and without LH68, at fixed $h=1/4$ and $h_\rho\to\infty$. 
Including LH68 with $\alpha=1$ slightly lowers the oscillation threshold $Re_c$ (from $\approx 4.5$ to $3$), consistent with the additional momentum flux from the small-scale wave spectrum. More strikingly, it strongly expands the range of periodic and quasi-periodic states: chaotic dynamics present 
above $Re\approx 75$ in P77 are suppressed up to $Re\approx 100$ for 
$\alpha=1$. Thus, adding small scale gravity wave forcing does not destabilize the system but 
instead restores regularity, extending the periodicity-resurgence 
results of \citet{leard_multimodal_2020} an d \citet{chartrand2024recovering} to the minimal LH68 
parameterization.

At $Re=100$, we observe in Fig.~\ref{fig:P77LH68bif} a transition from a chaotic state ($\alpha=0$) to a quasi-periodic state ($\alpha = 1$), with fast oscillations confined to the lower domain while the upper part 
approaches a period-one structure (panel~b). We interpret this as a 
competition between the two forcings: P77 dominates near the bottom 
where radiative attenuation is strongest, while LH68 — whose amplitude 
is altitude-independent — controls the upper domain. If $\alpha$ is further increased,  LH68 progressively wins this 
competition, driving the full profile toward periodic behavior.

\begin{figure}
\centering
\includegraphics[width=\columnwidth]{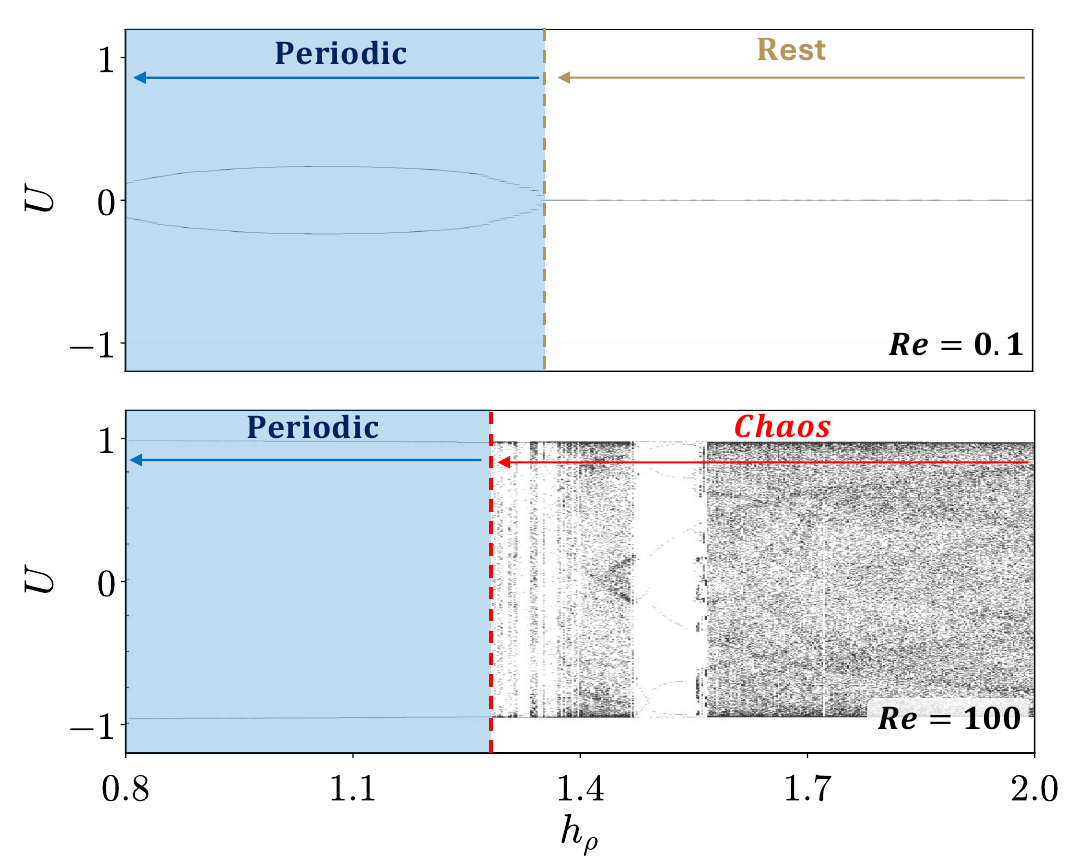}
\caption{Effect of exponential density decay (with $\alpha=0$). a) $Re=0.1$. b) $Re=100$.}
\label{fig:Density}
\end{figure}

Let us now turn to the the effect of exponential density variations.  Figure~\ref{fig:Density} shows that decreasing $h_\rho$ drives the system toward periodic regimes regardless of $Re$. In the low-Reynolds case (panel~a), exponential amplification of the forcing at small $h_\rho$ confines oscillations to the upper part of the domain. Panel~b shows an abrupt transition from chaotic to periodic dynamics below a critical value $h_\rho^c \approx 1.3$. No chaotic states are observed for $h_\rho < h_\rho^c$ up to $Re = 250$, indicating a stronger stabilization than that achieved by LH68 alone. This results 
from the exponential amplification of wave forcing at upper levels, 
which suppresses secondary extrema and accelerates convergence toward 
$|U|\to 1$, preventing interior shear layers from developing before 
they can seed chaos.

\textbf{Conclusion.} We have identified two distinct ways to restore periodic
oscillations in the Holton--Lindzen--Plumb system at large $Re$: (i) adding the
critical-level absorption of nearly undamped small-scale gravity waves, as
described by the LH68 parameterization; or (ii) accounting for the exponential
decay of the background density (P77$\rho$). With either effect, QBO-like
periodicity becomes much more robust and less dependent on small-scale viscosity
than in the P77 case. Result (i) highlights the importance of the vertical
attenuation length of the waves: weakly damped waves (with attenuation length
larger than the domain height), modeled either by spectra of P77-type waves, as in
\citep{leard_multimodal_2020,chartrand2024recovering}, or by LH68,
favor periodicity. We conjecture that this holds for other parameterizations of
weakly damped waves, including wave breaking below critical levels and saturation
\citep{campbell2005constraints2}. This remains to be established, as does the
extension to a turbulent viscosity enhanced where waves break
\citep{bremaud2026sensitivity}.

The competition between chaotic and periodic states can be summarized as follows.
On one side, stronger standing-wave forcing (high $Re$) destabilizes an
increasing number of high vertical modes, that is, vertical profiles of the zonal wind with multiple
extrema, whose amplification marks the transition to chaos
\citep{chartrand2024recovering}. On the other side, undamped-wave forcing (LH68)
and density stratification each counteract this growth. LH68 sweeps secondary
extrema toward the lower boundary through localized critical-level absorption
before they can grow. Exponential amplification of the forcing by density
variations instead favors the formation of a front aloft, as $|U|\to 1$, and
accelerates its descent before secondary extrema have time to develop. In both
cases, extrema are removed faster than they can grow, preventing chaos.

These mechanisms differ from standard strategies for taming chaos, which rely on feedback or weak periodic forcing applied from outside the system
\citep{ott1990controlling,braiman1991taming}. Here, all forcings are steady and periodicity emerges from wave--mean flow interactions only, which provides a route  complementary to synchronization with an imposed annual cycle \citep{rajendran2016synchronisation}. Considering such external forcing and generalizing the present framework to account for meridional circulation, either in a two-dimensional model \cite{plumb1982model} or through parameterized upwelling in one dimension \cite{saravanan_90,dunkerton_97,rajendran2016synchronisation,match_mean-flow_2020}, will be necessary to identify the dominant mechanisms governing the flow-reversal periodicity.

The question of what sets regularity versus chaos in a
wave-driven mean flow extends beyond QBO-like phenomena. On the one hand,
momentum deposition by critical-layer absorption of a broadband wave spectrum, as
in LH68, could play a role in equatorial deep oceanic jets
\citep{menesguen2019observations}, in the dynamics of multiple zonal jets in fluids and
plasmas \citep{galperin2019zonal,lemasquerier2025emergence}, and in wave--vortex interaction models
\citep{nolan2026hyper}. On the other hand, the momentum forcing in LH72 and P77 has inspired similar models for strato-rotational instabilities \cite{seelig2015can,meletti2026wave}, and is a streaming effect closely related to Faraday-wave-driven flows
\citep{higuera2008faraday} acoustic streaming \citep{launay2019transition}. Whether the mechanisms favoring periodic states
identified here have counterparts in these systems remains an open question.

\textbf{Acknowledgement}
We acknowledge Cerasela Calugaru and the whole team from PSMN/CBP at ENS de Lyon for computing facilities.  We warmly thank Ugo Martinez for his preliminary work on the numerical model, Louis Couston and Jason Reneuve for useful discussions on wind reversals with multiple wave forcing. This work was funded by French Research Agency (ANR) under BOURGEONS project, grant ANR-23-CE40-0014-01, SEEDLING project, grant  ANR-24-EXMA-0001, and CliMath project, grant ANR-23-EXMA-0003, PEPR Math-Vives. 
%% DATA AVAILABILITY STATEMENT
%\datastatement
%The model used for this study will be available on Zenodo. 

%\bibliographystyle{ametsocV6}
\bibliography{references}

\clearpage
\onecolumngrid        % if the main file is two-column
\appendix
\setcounter{equation}{0}
\setcounter{figure}{0}
\setcounter{table}{0}
\renewcommand{\theequation}{S\arabic{equation}}
\renewcommand{\thefigure}{S\arabic{figure}}
\renewcommand{\thetable}{S\arabic{table}}

\begin{center}
\textbf{\large Supplementary Material}
\end{center}

These supplementary materials contain
\begin{itemize}
    \item Appendix A:  Details on model derivation and adimensionalization.
    \item Appendix B: LH68 recovered from P77 with a continuum of weakly damped internal waves.
    \item Appendix C: Numerical details.
        \item Appendix D: Hovm\"{o}ller diagrams of QBO-like states with exponential density variations.
    \item  Appendix E: Adding back LH68 parameterization to the original Holton-Lindzen 72 model, including altitude-dependent wave attenuation length scale
\end{itemize}

\section{Appendix A:  Details on model derivation and adimensionalization}

The aim of this appendix is to show how the nondimensional  P77+LH68 model presented in the letter is related to the classical dimensional Holton-Lindzen-Plumb model augmented with Lindzen-Holton 68 parameterization.\\

\textbf{Quasi-linear framework.} The model assumes the dynamics is confined in a vertical plane above the equator, in a stratified fluid described by anelastic equations (sound waves are filtered out but strong variation of atmospheric density are accounted for). 
The starting point of the model is a decomposition between the zonally ($x$) averaged equatorial wind, \(\overline{u}(z,t)\) and its  fluctuations \((u',w')\), with no net vertical flow ($\overline{w}=0$). Owing to periodicity in $x$ direction, the zonally averaged horizontal momentum equation reads 
\begin{equation} \label{eq:HLP}
\frac{\partial \overline{u}}{\partial t} = \nu \frac{\partial^2 \overline{u}}{\partial z^2} 
- \frac{1}{\rho} \frac{\partial}{\partial z} \left( \rho \overline{u'w'}^p \right) 
- \frac{1}{\rho} \frac{\partial}{\partial z} \left( \rho \overline{u'w'}^{\text{sgw}} \right),
\end{equation}
where \(\nu\) is a constant  viscosity, and \(\rho(z) = \rho_0 \exp(-z/H_\rho)\) is the background atmospheric density with scale height \(H_\rho\). The total wave momentum flux $\overline{u'w'}$ has been separated into a \textit{planetary wave (p)} component \(\overline{u'w'}^p\) and a \textit{small-scale gravity wave (sgw)} component \(\overline{u'w'}^{\text{sgw}}\), neglecting (or time-filtering) cross-correlation terms. Importantly, the sgw component describes the effect of a continuum of small scale gravity waves, not just a single standing wave. Those quadratic  terms are computed using a quasilinear approach, assuming a time scale separation between mean flow evolution and the adjustement of the wave-field around this mean flow, and a scale separation between mean flow variations in $z$ direction and vertical wavelength, allowing for a WKB computation of the waves. More details on the derivation of this model and underlying hypotheses are given in \cite{renaud_holton-lindzen-plumb_2020}. In particular, the derivation relies on a distinguished limit of infinitely large Richardson number (large buoyancy frequency compared to the vertical shear of the horizontal mean flow). In this limit, wave momentum is fully absorbed at critical layers, and no shear instability occurs. In the real stratosphere, the Richardson number is generally large but can be locally small, potentially leading to wave over-reflection and instabilities. These effects are not accounted for in the present model; incorporating them is an interesting direction for future work.\\ 

\textbf{Holton-Lindzen 1972 (HL72) model for planetary waves.} In the original  formulation of the model \citep{holton_lindzen_72}, the planetary wave flux is modeled as a combination of equatorial Kelvin and Rossby waves:  
\begin{equation} \label{eq:HL72_flux}
\begin{aligned}
\rho \overline{u'w'}^p 
&= \rho_0 F_p \exp\!\left[-\!\int_0^z \frac{dz'}{\Lambda_K \left(1-\overline{u}(z')/c_K\right)^2}\right]  \\
&\quad - \rho_0  F_p \exp\!\left[-\!\int_0^z 
\frac{dz'}{\Lambda_R \left(1+\overline{u}(z')/c_R\right)^2}
\left(\frac{B}{1+\overline{u}(z')/c_R}-1\right)\right].
\end{aligned}
\end{equation}
where \(c_K\) and \(c_R\) are the horizontal phase speeds of the Kelvin and Rossby waves. Their vertical attenuation is controlled by radiative damping, with characteristic length scales
\begin{equation}
\Lambda_K = \frac{c_K^2 k_K}{N \gamma}, \quad \Lambda_R = \frac{c_R^2 k_R}{N \gamma},
\end{equation}
depending on the buoyancy frequency \(N\), damping rate \(\gamma\), and horizontal wavenumbers \(k_K\) and \(k_R\). Rotation of the planet is controlled by $B = \beta/ k_R^2 c_R$ with $\beta$ the gradient of Coriolis parameter at the equator.\\  

\textbf{Plumb 1977 (P77) simplification.} 
The special case \(B=0\), \(c_K = c_R=c\), and \(\Lambda_R = \Lambda_K=\Lambda\) corresponds to Plumb (1977), and involves a single standing internal gravity wave forcing. The model can also be generalized to multiple wave forcings, where each wave is associated with a pair \((c,\Lambda)\), or equivalently with a given frequency and wavenumber. Each wave then contributes independently of the others \cite{saravanan_90}.\\

\textbf{LH68 parameterization.} The momentum flux induced by a continuum of small scale gravity waves is modelled using LH68 formulation. \cite{lindzen_holton_68} built on  \cite{brooker_bretherton_1967} to describe how undamped small-scale gravity waves deposit their momentum at the lowest critical level, where their phase speed coincides with the mean flow velocity:  
\begin{equation} \label{eq:LH68param}
-\frac{1}{\rho} \frac{\partial}{\partial z} \left( \rho \overline{u'w'}^{\text{sgw}} \right) = 
\begin{cases}
f(\overline{u}) \frac{\partial \overline{u}}{\partial z}, & z = \min \{ z' \,|\, \overline{u}(z',t) = \overline{u}(z,t) \}, \\
0, & \text{otherwise}.
\end{cases}
\end{equation}
Here, \(f(c)\) is the momentum flux density at the bottom of the stratosphere, integrated over all phase speeds:  
\begin{equation}
\rho \overline{u'w'}^{\text{sgw}} \big|_{z=0} = \int_{c_{\min}}^{c_{\max}} \text{sgn}(c) f(c) \, dc,
\end{equation}
with \(\text{sgn}(c)\) the sign function. The LH68 parameterization transfers momentum to the mean flow once a wave encounters their lowest critical level.

In presence of multiple critical levels in a given profile, only the lowest one is subject to momentum forcing. Some part of the wind profile are therefore not subject to any forcing. This concerns also altitudes where the wind velocity exceeds the range of  horizontal phase speed in the forcing.\\

\textbf{Adimensionalization.} We now consider the P77 setting of HLP, with $B=0$, $c_K=c_R \equiv c$,  $\Lambda_K=\Lambda_R \equiv \Lambda$ in Eq.~(\ref{eq:HL72_flux}). 

We assume that small-scale gravity wave momentum flux at the bottom is described by a water-bag density distribution with constant amplitude, i.e., 
\begin{equation}
    f(c') = \frac{F_{sgw}}{c} \text{ for } c' \in [-c, c],
\end{equation} and $f(c')=0$ elsewhere.
This simple waterbag momentum-forcing distribution is sufficient to reproduce the generic behavior obtained for a broader range of wave-forcing distributions in \cite{leard_multimodal_2020}. Note also that results are unchanged for a waterbag distribution where for $f(c')=F_{sgw}/c$ for $c^\prime\in[-B,B]$, when $B>c$.

Let us recall for completeness that the system admits four non-dimensional parameters 
\begin{equation}
    h = \frac{H}{\Lambda}, \quad Re = \frac{F_p \Lambda}{\nu c}, \quad
    h_\rho = \frac{H_\rho}{\Lambda}, \quad \alpha = \frac{F_{sgw}}{F_p},
\end{equation}
with $F_p$ the bottom amplitude of the momentum flux per unit mass for each of the planetary waves.  Following \cite{vallis2017atmospheric}, we adimensionalize the model using
\begin{equation}
    U = \frac{\overline{u}}{c}, \quad Z = \frac{z}{\Lambda}, \quad T = \frac{t F_p}{c \Lambda},
\end{equation}
which leads to the rescaled equation and boundary conditions:
\begin{equation} \label{eq:extendedHLP_dimensionless}
\frac{\partial U}{\partial T} = \frac{1}{Re} \frac{\partial^2 U}{\partial Z^2} - e^{Z/h_\rho} \frac{\partial}{\partial Z} \Big( \mathcal{F}_p[U] +  \mathcal{F}_{sgw}[U] \Big), 
\end{equation}
\begin{equation}
  \mathcal{F}_p[U] = \exp\left[-\int_0^Z \frac{dZ'}{(1-U)^2}-\right]-\exp\left[-\int_0^Z \frac{dZ'}{(1+U)^2}-\right],\quad   \mathcal{F}_{sgw}[U]=-\alpha \Phi[U]
\end{equation}
with $\Phi[U]=0$ whenever there exists a depth $Z'<Z$ such that $U(Z',T) = U(Z,T)$, and $\Phi[U]=U$ otherwise, consistently with the LH68 critical-level formulation with waterbag forcing.  The system is complemented with boundary conditions 
\begin{equation}
   U|_{Z=0} = 0,  \qquad
    \frac{\partial U}{\partial Z} \bigg|_{Z = h}= 0 , 
\end{equation}

These equations provide a minimal model capturing the combined effects of planetary waves (controlled by $Re$) and small-scale gravity waves (controlled by $\alpha Re$).  

The textbook model for the QBO explanation is recovered in the limits $h_\rho \to \infty$ (Boussinesq atmosphere), $h \to +\infty$ (semi-infinite stratosphere), and $\alpha=0$ (planetary waves only). In that case, the first bifurcation from a rest state to a limit cycle as $Re$ is increased was described by \cite{plumb_77} and studied in depth by \cite{semin_nonlinear_2018}. The emergence of other oscillatory regimes as $Re$ is further increased has been explored by \cite{renaud_periodicity_2019}, while the impact of changes in the effective stratospheric depth $h$ on both the first bifurcation and chaotic states was addressed by \cite{chartrand2024recovering}. The influence of the density scale height $h_\rho$ on the periodic regime was described first by \cite{plumb_77}. The role of the small-scale gravity wave parameter $\alpha$ in promoting transitions toward periodic states was discussed by \cite{dunkerton_97}. This effect has also been addressed by \cite{campbell2005constraints1, campbell2005constraints2} for more general small-scale gravity-wave parameterizations. Finally, \cite{match_mean-flow_2020} examined how $\alpha$ shapes the QBO buffer layer. In the present study, we  show that increasing $\alpha$ and decreasing $h_\rho$ suppress chaotic states, thereby enhancing the robustness of regular QBO-like oscillations.

\begin{figure*}
\centering
\includegraphics[width=\textwidth]{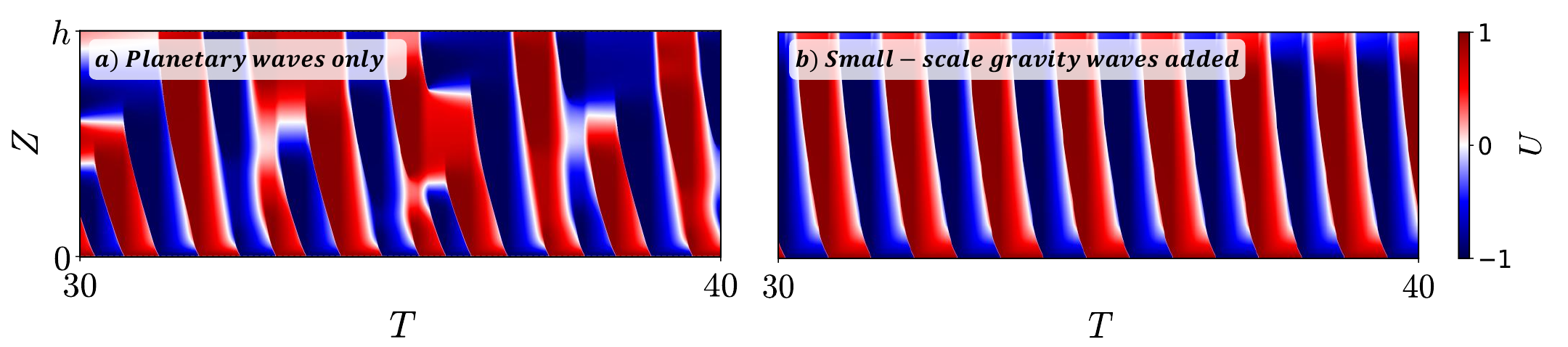}
\caption{Simulations of Holton-Lindzen-Plumb model for horizontal mean flow reversals identical to figure 1 of the main text except for the lower boundary condition $\frac{\partial U}{\partial Z} \bigg|_{Z = 0}= 0$.}
\label{fig:No_Slip}
\end{figure*}

We show in figure \ref{fig:No_Slip} that changing the lower boundary condition from no-slip  to free-slip does not modify qualitatively the periodicity recovery described in figure 1 of the main text. The fact that free-slip and no-slip boundary conditions does not change the system behavior at a qualitative level is  consistent with the results of \cite{rajendran2016synchronisation} when adressing QBO  synchronisation  with seasonal forcing and  \cite{renaud_holton-lindzen-plumb_2020} on the first bifurcation and subsequent quasi-periodic route to chaos.

\section{Appendix B: LH68 recovered from P77 with a continuum of weakly damped internal waves.}

We show in this appendix that LH68 can be interpreted as a mean-field model for a weakly attenuated continuum of  waves described by Plumb 77 parameterization. Let us consider a continuum of internal waves spanning a range of phase speeds \(c \in [c_\text{min},c_\text{max}]\) and fixed wavenumber \(k\) within the HLP framework. Using notation from the main text, the small-scale gravity wave momentum flux can be expressed as
\begin{equation}
  \rho  \overline{u'w'}^{sgw} = \int_{c_\text{min}}^{c_\text{max}} \operatorname{sgn}(c) f(c) 
  \exp\left(-\int_0^z \frac{dz'}{\Lambda_k(c) \left(1-\frac{\overline{u}(z')}{c}\right)^2} \right) dc,
\end{equation}
where \(\Lambda_k(c)=c^2 k / N \gamma\) is the attenuation length of each wave and \(f(c)\) is the momentum flux density at the base of the stratosphere.

\begin{figure*}[h!]
\centering
\includegraphics[width=0.7\textwidth]{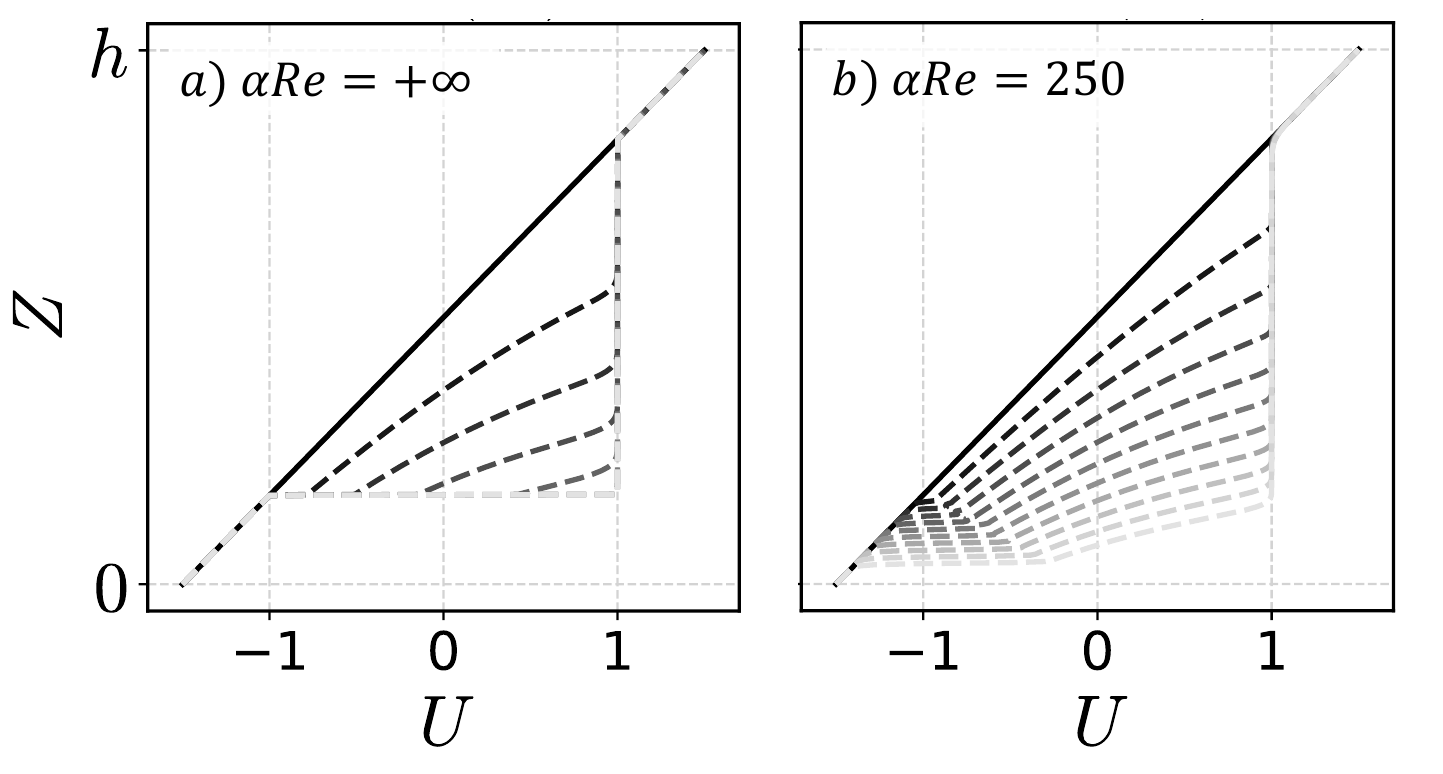}
\caption{Temporal evolution of an initially linear wind profile (plain black line), forced by the LH68 parameterization with a waterbag distribution $f(U)$ of undamped waves at the bottom boundary ($f(U)$ constant for $-1\le U\le 1$, zero otherwise).  a) Without explicit viscosity (a configuration similar to that of Sec.~4 of LH68) b) With a weak viscosity, set by $\alpha Re=250$ (and $Re\to 0$, since there is no standing-wave forcing). In both panels, $h_\rho=1.4$.}
\label{fig:LH68evolution}
\end{figure*}

Assuming that the wavenumber \(k\) of the small-scale waves is much larger than that of planetary waves, consistent with generation by localized convection, we consider the limit \(k\to \infty\). Physically, this corresponds to weakly damped waves that propagate nearly freely until they reach their first critical level, with \(\Lambda_k \gg H_\text{strat}\) (stratosphere height). In this limit, the exponential term approaches a Heaviside function:

\begin{equation}
    \lim_{\Lambda_k \to \infty} 
    \exp\left(-\frac{1}{\Lambda_k}\int_0^z \frac{c^2 dz'}{(c-\overline{u})^2}\right)
    =
    \begin{cases}
        1 & z < z_c,\\
        0 & z \ge z_c,
    \end{cases}
\end{equation}
where \(z_c\) is the lowest altitude where the critical level condition \(\overline{u}(z_c)=c\) is satisfied. Using this, the total momentum flux becomes
\begin{equation}
  \rho \overline{u'w'}^{sgw} = \int_{c_\text{min}}^{c_\text{max}} \operatorname{sgn}(c) f(c)\, \mathcal{H}(z_c - z) \, dc,
\end{equation}
where \(\mathcal{H}\) is the Heaviside step function. Taking the vertical derivative and applying the identity
\(\int f(x) \delta(g(x)) dx = \sum_i f(x_i)/|g'(x_i)|\) for \(x_i\) such that \(g(x_i)=0\), we obtain:
\begin{equation}
- \frac{\partial}{\partial z} (\rho \overline{u'w'}^{sgw}) = \operatorname{sgn}(\overline{u}(z)) f(\overline{u}(z)) \left|\frac{\partial \overline{u}(z)}{\partial z}\right|.
\end{equation}
This expression applies to the first critical level encountered by each wave; at higher critical levels, all momentum has already been absorbed, so the flux divergence is zero. With the no-slip boundary condition \(\overline{u}(z=0)=0\), we have \(\operatorname{sgn}(\partial_z \overline{u}) = \operatorname{sgn}(\overline{u})\) at altitudes where LH68 is active, recovering the original parameterization (Eq.~\ref{eq:LH68param}). This allows us to safely use LH68 to interpret the role of weakly damped waves described by \citep{leard_multimodal_2020,chartrand2024recovering} within the HLP framework. 
\\
\\
To validate our implementation of the LH68 parameterization, we show in Figure \ref{fig:LH68evolution} a configuration similar to Section 4 of the original Lindzen and Holton (1968) paper \cite{lindzen_holton_68}. Panel (a) presents a case without viscosity, as in Figure 2 of \cite{lindzen_holton_68}, while panel (b) presents a case with the same viscosity as in the main text (our Figure 2).
%
%Vraiment chouette la figure !!
%

\section{Appendix C: Numerical details \label{app:numerics}}
We solve (\ref{eq:extendedHLP_dimensionless}) using a 
finite difference semi-implicit scheme: the viscous term 
is treated implicitly, while the nonlinear advection term  is advanced using a first-order upwind scheme. Higher-order  schemes were tested but found to produce spurious oscillations near the front where $U \to 1$, which destabilize the simulation since any overshoot violates the physical bound $|U| < 1$.

Instabilities also arise when the vertical resolution is 
too coarse. The WKB integral in the wave forcing term 
becomes poorly resolved near the critical level, leading 
to spurious overestimation of the forcing just above it 
and to inaccurate localization of the critical-level 
height. Both effects produce strong fluctuations in the 
forcing near the critical level.

To determine empirically compatible values of $\Delta T$ and $\Delta Z$ ensuring both CFL condition and convergence of the WKB wave forcing estimate, we defined a critical time step as follows: for given parameters $(h, Re)$ and given $\Delta Z$, starting from an initial profile $U(Z)$  that reaches $U = 1$ at a known altitude, we advance the mean flow by one explicit time step and identify the 
largest $\Delta T$ for which the bound $|U| < 1$ is 
not violated at or below the front. Because of the previous effects, this critical time step becomes strongly dependent on the vertical resolution when the grid is too coarse, and converges as $\Delta z$ goes to 0. We consider that the integral is well resolved when this critical time step becomes constant with respect to the grid size

This provides a 
physically grounded criterion that is more 
stringent than a standard CFL condition. With this criterion, it was not necessary to use  the
ad hoc threshold on the maximum allowed velocity that was needed in \cite{renaud_periodicity_2019,leard_multimodal_2020,chartrand2024recovering}. 

Based on this procedure, we adopted $\Delta Z = 1/500$ as a conservative choice: at 
this resolution, the critical time step varies smoothly 
with $h$ and $Re$, 
indicating that the WKB integral is well-resolved across 
the parameter range explored in this study. The time step 
$\Delta t$ is then set just below the critical value for 
each parameter combination. 

\begin{figure*}
\centering
\includegraphics[width=\textwidth]{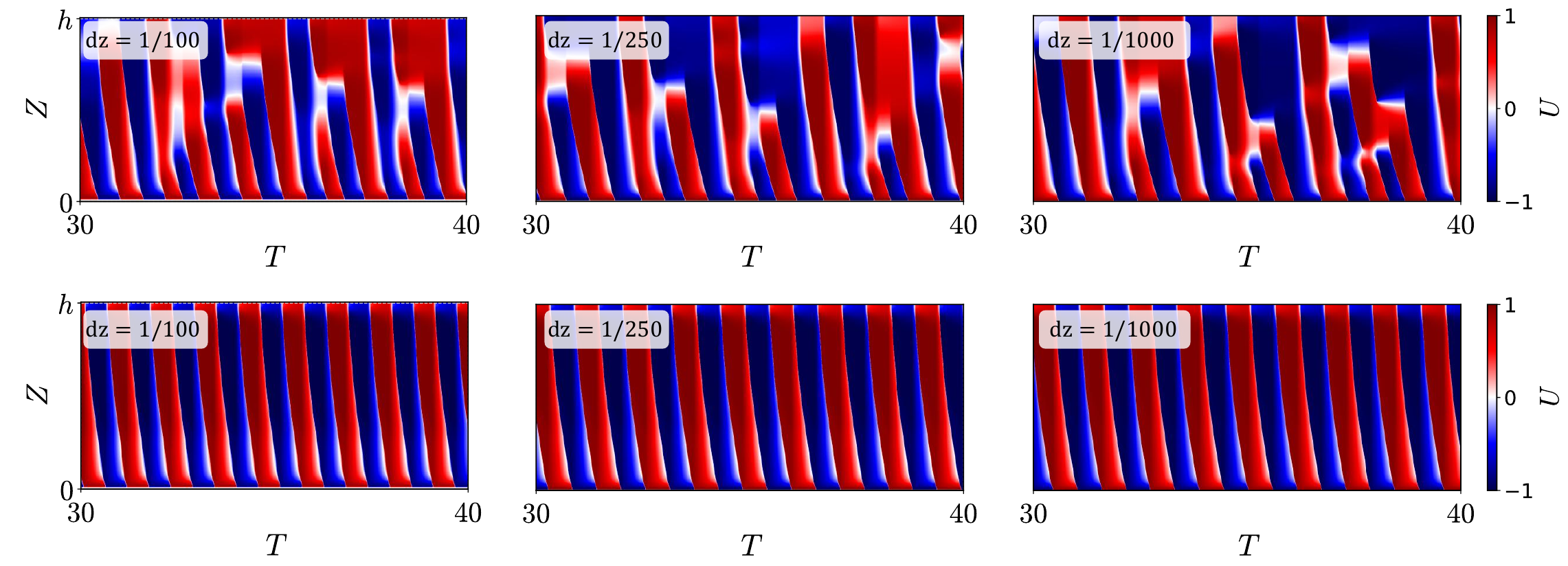}
\caption{Hovmöller diagrams illustrating the effect of resolution the periodicity recovery. Parameters are the same as in figure 1 of the main text, only the number of vertical levels varies. Figure 1 was done using $dz=1/500$.}
\label{fig:Resolution}
\end{figure*}

The sensitivity of the HLP model to numerical resolution was studied by \cite{renaud_periodicity_2019}, which demonstrated the robustness of the quasi-periodic route to chaos. Figure~\ref{fig:Resolution} further illustrates the robustness of periodicity recovery through exponential density stratification and small-scale gravity waves parameterized with LH68. Note that in the chaotic regime, the solution behavior is robust to changes in vertical resolution, though the solution itself is not, as expected due to sensitivity to initial conditions.

We also analyze the sensitivity of the LH68 parameterization to vertical resolution by comparing the temporal evolution of the profile shown in Figure~2 of the main text for different grid spacings. The comparison between panel (b) of Figure~2 in the main text and Figure~3 of this supplementary material demonstrates that our vertical resolution is sufficient to ensure that the LH68 dynamics are well-resolved.

\begin{figure*}
\centering
\includegraphics[width=\textwidth]{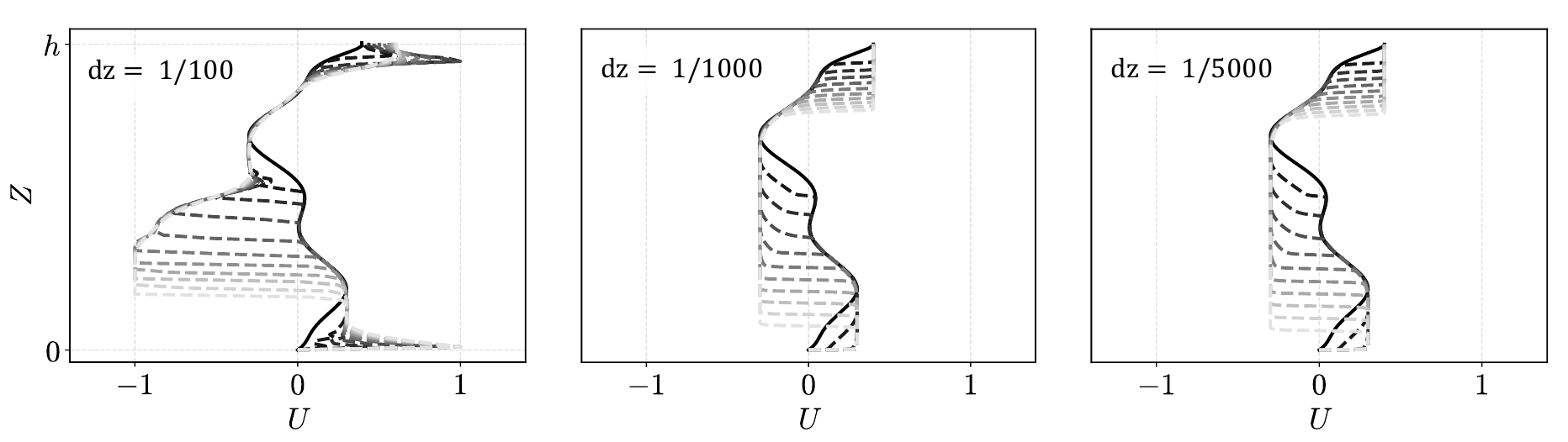}
\caption{Temporal evolution of a wind profile (initial state : plain black line) for different vertical resolutions. Parameters are the same as in panel b) of figure 2 from the main text, only the number of vertical levels varies. Figure 2 was done using $dz=1/500$.}
\label{fig:sensitivityLH68}
\end{figure*}

\section{Appendix D: Hovm\"{o}ller diagrams of QBO-like states with exponential density variations}

\begin{figure*}
\centering
\includegraphics[width=\textwidth]{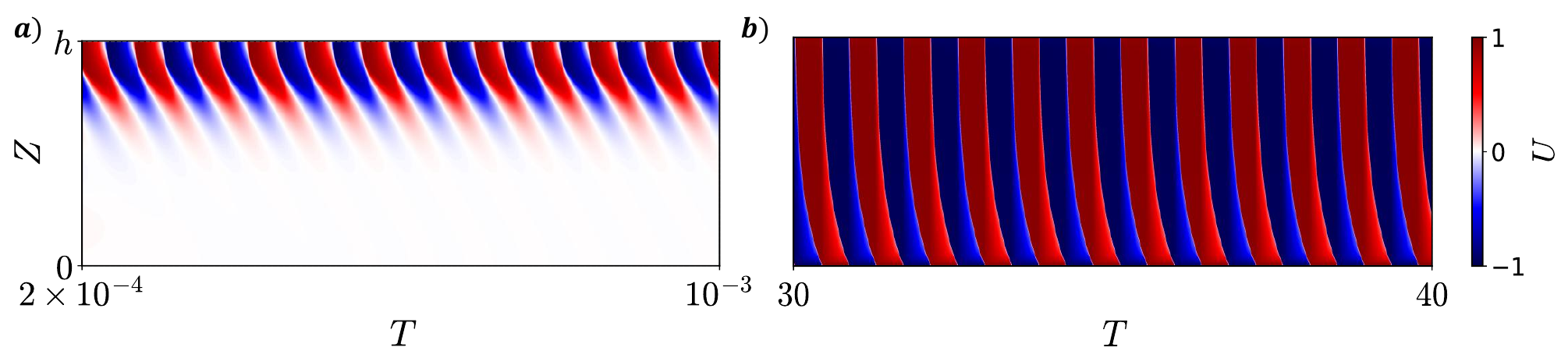}
\caption{Hovmöller diagrams a) $(Re,h,h_{\rho},\alpha)=(10^{-3},1/4,0.28,0)$} b) $(Re,h,h_{\rho},\alpha)=(100,1/4,0.9,0)$
\label{fig:Density_Appendix}
\end{figure*}

Figure~\ref{fig:Density_Appendix} shows typical QBO-like states obtained with the P77$\rho$ configuration, i.e. planetary-wave forcing with exponential background density variations. In panel~a, weak forcing combined with exponential density decay leads to intensified dynamics near the top of the domain, so that the oscillation threshold is exceeded only at high altitude. Note the qualitative similarity with the observed pattern of the Semi-Annual Oscillation in the upper stratosphere. Panel~b illustrates how a state that is chaotic in the uniform-density limit ($h_\rho \to \infty$) recovers periodicity when $h_\rho$ is sufficiently small.

\section{Appendix E: adding LH68 back to HL72, including altitude-dependent wave attenuation length scale} 

We now show that the results obtained with P77 model holds for the more general HL72 model, where the two 
counter-propagating waves driving the QBO-like oscillation 
have distinct propagation characteristics. This asymmetry 
is a direct consequence of Earth's rotation on the 
equatorial waveguide: the two waves correspond to an 
equatorially trapped Kelvin wave and a Rossby wave, which 
differ in both their dispersion relations and attenuation 
length scales. 

The HL72 framework is of particular interest here for two 
reasons. First, it is historically the reference model for 
the QBO mechanism. Second, and more relevantly for the 
present study, it provides the natural setting in which to examine the combined effect of planetary and small-scale internal waves.

Concretely, the asymmetry is captured by the parameter 
$B$ appearing in the Rossby-wave momentum flux term in 
Eq.~(\ref{eq:HL72_flux}), and by the difference between the attenuation length scales $\Lambda_K$ and $\Lambda_R$ 
of the Kelvin and Rossby waves respectively. Following HL72 we consider $c_K/c_R=1$, $\Lambda_K/\Lambda_R=1/4$ and $B=1.92$. As a result 
of this symmetry breaking, the zero-wind state is no 
longer a steady solution of the system, and the 
transition to a QBO-like oscillation is more complex than 
in the P77 case. Nevertheless, a quasi-periodic route to 
chaos persists \citep{renaud_periodicity_2019}, and the 
insights gained from the symmetric case carry over, as 
we discuss below.

HL72 also emphasized the need for an increasing damping rate with altitude to reproduce a realistic QBO structure. We adopt the same formulation here, defining : \begin{equation}
    \gamma(Z)=
\begin{cases} \label{eq : varying attenuation lengthscale}
\frac{1}{a}+(\frac{1}{b}-\frac{1}{a})\frac{Z}{Z_V} & \text{if}~~ Z \le Z_V \\
\frac{1}{b} & \text{otherwise} 
\end{cases}
\end{equation}
with $a=1.8144 \times 10^6$, $b=6.048 \times 10^5$ and $Z_V \approx 2.9$. In that case the attenuation length scale $\Lambda(Z)=\omega^2/(Nk\gamma)$ is a function of altitude. The importance of this altitude-dependent parameter in shaping the vertical structure of the QBO  on Earth has been emphasized by \citep{hamilton1981vertical}. Based on the analysis performed in the main text, we expect that this increase of wave attenuation would counteract the effect of exponential density variations and favor chaotic over periodic regimes at large $Re$, as we verify in the following.

\begin{figure*}
\centering
\includegraphics[width=\textwidth]{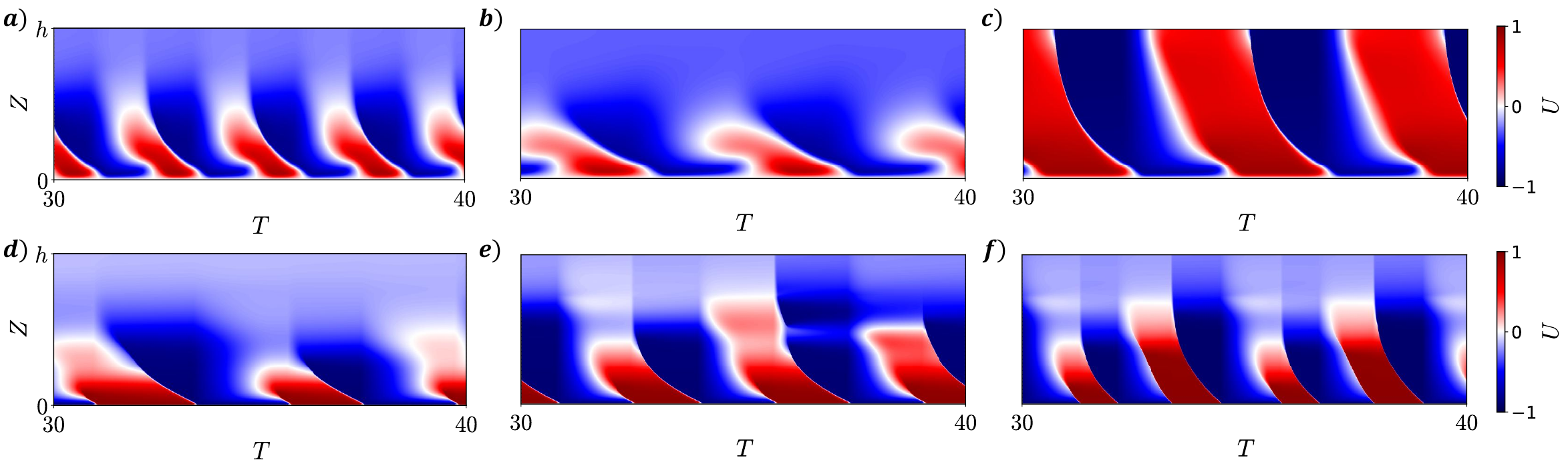}
\caption{Hovmöller diagrams illustrating the dynamics of the modified HL72 model a) Reference HL72 simulation $(Re,h,h_{\rho},\alpha)=(8,4,1.4,0)$ with varying attenuation coefficient $\gamma(Z)$ (Eq. \ref{eq : varying attenuation lengthscale}) b) Reference HL72 simulation with $h_{\rho}=+ \infty$ c) Reference HL72 simulation with constant attenuation lengthscale $\Lambda_K$ d) $(Re,h,h_{\rho},\alpha)=(70,4,+\infty,0)$ e) $(Re,h,h_{\rho},\alpha)=(70,4,1.4,0)$ f) $(Re,h,h_{\rho},\alpha)=(70,4,1.4,1) $}
\label{fig:HL72}
\end{figure*}

 Figure~\ref{fig:HL72} illustrates different dynamical regimes in this setup. 
Non-dimensional parameters $Re$, $h$ and $h_{\rho}$ are obtained by considering the value of $\Lambda$ at $z=0$. Panel a reproduces the original HL72 parameters, yielding periodic oscillations. This periodic behavior persists both when exponential density stratification is removed (panel b) and when increased wave attenuation is removed (panel c, with $\Lambda=\Lambda(0)$ held fixed), highlighting the key role of enhanced wave attenuation at high altitude in producing QBO-like oscillations. Indeed, this enhanced attenuation compensates for the exponential growth of wave forcing with altitude.

So far, we have assumed a depth-independent radiative attenuation. More realistic models of the stratosphere instead let the attenuation increase with altitude \citep{holton_lindzen_72}, a feature that plays a key role in shaping the vertical structure of the observed QBO on Earth \citep{hamilton1981vertical}. Based on our analysis, we expect that this increase of wave attenuation would counteract the effect of exponential density variations and favor chaotic over periodic regimes at large $Re$, as we verified in the original \citet{holton_lindzen_72} setting (see Supplementary Material).

Panels d and e show quasi-periodic and chaotic states obtained by increasing the Reynolds number relative to the original HL72 study, with and without exponential density stratification respectively. Panel f demonstrates that adding the LH68 parameterization restores order: the chaotic state of panel e transitions to a quasi-periodic regime.

\bibliography{references}

\end{document}